\documentclass[reprint,amssymb,pra]{revtex4-1}
\usepackage[utf8]{inputenc}
\usepackage[T1]{fontenc}
\usepackage{graphicx}
\usepackage{amsmath}
\usepackage{amssymb}
\usepackage{mathrsfs}
\usepackage{braket}
\usepackage{geometry}
\usepackage{here}
\usepackage{lmodern}
\usepackage{array}
\usepackage{natbib}
\usepackage{url}
\usepackage{textcase}
\usepackage{bm}
\usepackage{natbib}
\usepackage{color}
\begin{document}
	\title{Laser control of ultracold molecule formation: The case of RbSr}
	
\author{Adrien Devolder}
\altaffiliation{Laboratoire Aimé Cotton, CNRS, Université Paris-Sud, ENS Paris-Saclay, Université Paris-Saclay, B\^at. 505, 91405 Orsay Cedex, France }
\author{Mich\`ele Desouter-Lecomte}
\altaffiliation{Institut de Chimie Physique, CNRS, Université Paris-Sud, Université Paris-Saclay, Bât. 349,  91405, Orsay Cedex, France}
\author{Osman Atabek}
\altaffiliation{Institut des Sciences Moléculaires d'Orsay, CNRS, Université Paris-Sud, Université Paris-Saclay, Bât. 520, 91405 Orsay Cedex, France }
\author{Eliane Luc-Koenig}
\altaffiliation{Laboratoire Aimé Cotton, CNRS, Université Paris-Sud, ENS Paris-Saclay, Université Paris-Saclay, B\^at. 505, 91405 Orsay Cedex, France }
\author{Olivier Dulieu}
\altaffiliation{Laboratoire Aimé Cotton, CNRS, Université Paris-Sud, ENS Paris-Saclay, Université Paris-Saclay, B\^at. 505, 91405 Orsay Cedex, France }

\begin{abstract}
We have studied the formation of ultracold RbSr molecules with laser pulses. After discussing the advantages of the Mott insulator phase for the control with pulses, we present two classes of strategies. The first class involves two electronic states. Two extensions of stimulated Raman adiabatic passage (STIRAP) for multi-level transitions are used : alternating STIRAP (A-STIRAP) and straddle STIRAP (S-STIRAP). Both transfer dynamics are modeled and compared. The second class of strategies  involves only the electronic ground state and uses infrared (IR)/TeraHertz (THz) pulses. The chemical bond is first created by the application of a THz chirped pulse or  $\pi$-pulse. Subsequently, the molecules are transferred to their ro-vibrational ground state using IR pulses. For this last step, different optimized pulse sequences through optimal control techniques, have been studied. The relative merits of these strategies in terms of efficiency and robustness are discussed within the experimental feasibility criteria of present laser technology.
\end{abstract}

\maketitle
	
	\section{Introduction}
Ultracold molecules are promising systems for a large number of applications, including quantum simulation \cite{quant_simu_1,quant_simu_2,quant_simu_3,Cold_molecule_Simu_1,Cold_molecule_Simu_2}, quantum computation \cite{Quantum_comp_mol,RbSr_quant_comp_1,RbSr_quant_comp_2,Cold_molecule_Inf_1,Cold_molecule_Inf_2,Cold_molecule_Inf_3,
Sugny2009,Bomble2010,Jaouadi2013,Friedrich2013}, ultracold chemistry \cite{Ultra_chem,Kotochigova2015} and precision measurements \cite{Cold_molecule_const_1,Cold_molecule_const_2,edm_1,edm_2,Cold_molecule_PT}. Therefore, their formation remains an important objective. Ultracold alkali-alkali molecules have been first to be created. For that purpose, the most commonly used method is  magnetoassociation \cite{magnetoassociation_2,magnetoassociation_7} based on a Magnetic Feshbach Resonance (MFR) \cite{Magnetic_Feshbach_resonance_1,Magnetic_Feshbach_resonance_2,Magnetic_Feshbach_resonance_3}. After their formation the molecules occupy a loosely-bound state, not suited for the above-cited applications. A transfer to the absolute ground-state is necessary and made via a STImulated Raman Adiabatic Passage (STIRAP) \cite{relax_STIRAP_3,Liu2019}. 
Magnetoassociation still remains a challenge for the formation of molecules involving a closed-shell atom, like alkaline-earth or Ytterbium. This is precisely the case of RbSr which is a promising candidate for building a quantum simulator of lattice spin systems \cite{quant_simu_2}, one of the main Hamiltonian model in condensed matter physics. Recently, narrow MFRs for RbSr were observed in F. Schreck's group \cite{RbSr_15} but not yet exploited for magnetoassociation. 
\begin{figure}
	\centering
	\includegraphics[width=8cm]{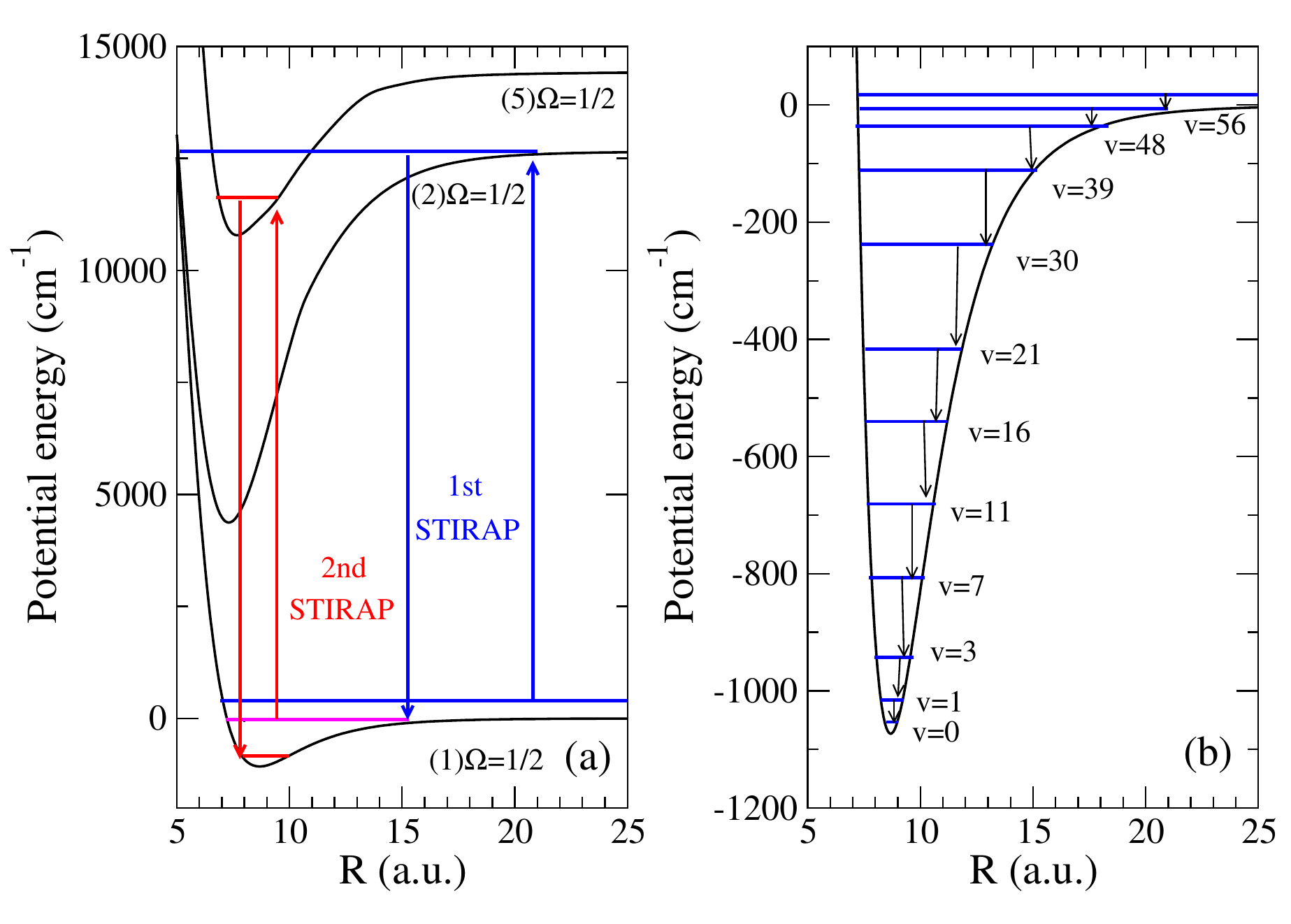}
	\caption{Two strategies for creating RbSr molecules in their absolute ground state, either via two STIRAP transfers (left panel), or using THz/infrared pulses  inducing rovibrational transitions inside the electronic ground state. In the left panel, the levels involved in the first STIRAP are drawn in blue, while the ones involved in the second STIRAP are drawn in red. The intermediate level, involved in both schemes, is drawn in magenta. In the right panel, the vibrational levels involved in the transfer are drawn as  blue lines. }
	\label{fig:strategy}
\end{figure}

An alternative to magnetic methods is the formation using lasers. Actually, the first ultracold molecules were created with continuous-wave (cw) lasers via photoassociation \cite{Photoassociation_first}. The efficiency of photoassociation is however limited by atom losses, induced by spontaneous emission.  For this reason, photoassociation has temporarily been replaced by magnetoassociation for the formation of alkali-alkali molecules. The above mentionned limitations of magnetoassociation when addressing closed-shell atoms offer a new perspective for laser methods, especially if spontaneous emission can be avoided, which can hardly be done when referring to cw lasers, that will additionally lead to Rabi oscillations.  For an efficient formation, a time-dependence must be introduced through the use of pulsed lasers. It is worthwhile noting that, a time-dependence is also present in magnetoassociation, allowing the adiabatic passage. 

 In this paper, we consider two different strategies (see Figure \ref{fig:strategy}). The first is STIRAP, which was already used for the formation of ultracold Sr$_2$ molecules \cite{STIRAP_tight_trap_1,STIRAP_tight_trap_2} and implies excited electronic states during the transfer (Figure\ref{fig:strategy}(a)). More particularly, advanced derivatives of STIRAP were considered, A-STIRAP and S-STIRAP \cite{Tannor97,Tannor99,Vitanov2017}. The second strategy addresses only the electronic ground state via the use of Infrared/Terahertz lasers (Figure \ref{fig:strategy}(b)). This last approach was previously suggested by Kotochigova \cite{LASIFR_1} and Juarros $et$ $al.$ \cite{LASIFR_2} referring to cw lasers. 

The structure of the paper is the following. In Section II, we discuss the issues related to the use of pulsed lasers for the formation of molecules and how we can circumvent them. In Section III, the Hamiltonian of atomic pairs in Mott insulators is presented. The formation of molecules following the first strategy via STIRAP methods is discussed in Section IV. Section V is devoted to the second strategy involving a single electronic state. Finally, we conclude in Section VI. 

\section{ Formation of ultracold molecules with pulsed lasers}
Different issues must be solved when attempting the formation of a chemical bond with laser pulses. First, pulses only interact with colliding atom pairs. Unfortunately, this involves only a few atoms. Then, a small number of molecules are created during a single pulse. For the conversion of all atom pairs to ultracold molecules, many successive applications of the same pulses are required. However, as shown for pulsed photoassociation \cite{Luc-Koenig2004}, the application of pulses can also dissociate the molecules previously formed, unless control strategies aiming at the formation of very long-lived Feshbach resonances  (Zero Width Resonances) are specifically addressed \cite{ZWR}. The second caveat is that the colliding pairs do not occupy a single translational state but a distribution of translational states. As the coherent control with laser pulses is an unitary process and cannot reduce the entropy, the occupation of a single state at the end of the process is unlikely. 

These difficulties can be solved by trapping ultracold atoms in an optical lattice. By increasing the intensity of the trapping lasers, a particular state of matter can be achieved : a Mott insulator \cite{Book_optical_lattice}, where the tunneling between the sites is suppressed. As a consequence, the number of atoms in each site can be controlled. Attempting the formation of homonuclear diatomic molecules, a Mott insulator with two atoms per site is the starting point \cite{Volz2006,Thalhammer2006}, while for heteronuclear diatomic molecules, the starting point is the overlap of two Mott insulators of one atom per site (one for each species)\cite{Damski2003,Stoferle2006,Ospelkaus2006,Freericks2010}. In both cases, a lattice of atom pairs is obtained and these pairs are well isolated from each other. 
The trapping in Mott insulators can solve the two above-cited issues. First, the pulse sequence can be applied to all pairs at the same time and independently. Therefore, merely a single pulse sequence is enough. Secondly, the trapping induces the quantization of translational motions. If the energy gap between the quantized translational states is higher than the thermal energy,  every pair occupies the lowest translation state. The entropy is then minimal. The control of a scattering problem becomes the transfer between discretized states, for which some efficient methods exist, such as the adiabatic passage with a chirped pulse, the $\pi$-pulse or the STIRAP \cite{Vitanov2001}. 

An alternative to the trapping in Mott insulators is the use of optical tweezers. In this case, one pair of atoms is trapped by optical tweezers and we can control the formation of a single molecule. This has been accomplished in the group of K.-K. Ni \cite{Optical_Tweezer_1,Optical_Tweezer_2}. In the following, we focus on the Mott insulator, all discussed strategies presenting however the potentiality to be transposed to the case of trapping by optical tweezers.

\section{Hamiltonian of a trapped atomic pair}
\subsection{Separation of the center-of-mass  and the relative motion}
The following situation is under consideration: 
Two Mott insulators with respectively an  $^{87}$Rb and an $^{84}$Sr per site are overlapping. In each site, the Rb/Sr atoms  (of mass $m_{Rb}/m_{Sr}$) feel a harmonic potential of frequency $\omega_{Rb}$/$\omega_{Sr}$. The total Hamiltonian for the center-of-mass $\vec{R}_{CM}$ and the relative  $\vec{R}$ motions of the atomic pair is given by \cite{Optical_latt_trap} :
\begin{equation}
\begin{split}
\hat{H}_{\textrm{trap}}=&-\frac{\hbar^2}{2M}\nabla_{\vec{R}_\textrm{CM}}^2+\frac{1}{2} M \omega^2_{\textrm{CM}} R_{\textrm{CM}}^2-\frac{\hbar^2}{2\mu}\nabla_{R}^2 +\hat{H}_{rot}\\ &+\hat{H}_{el}+ \hat{H}_{SO}+\frac{1}{2}\mu\omega^2_{rel}R^2 +\mu\Delta\omega\vec{R}_{\textrm{CM}}.\vec{R}
\end{split}.
\label{eq:Htrap}
\end{equation}
The first term corresponds to the kinetic energy of the center-of-mass with total mass $M$. The second term is the harmonic potential for the center-of-mass motion in the trap with the frequency $\omega_{\textrm{CM}}=\sqrt{\frac{m_{Rb}\omega_{Rb}^2+m_{Sr}\omega_{Sr}^2}{m_{Rb}+m_{Sr}}}$. The third one is the kinetic energy of the relative motion. $\hat{H}_{rot}=\frac{\hbar^2\ell^2}{2\mu R^2}$ is the relative rotational motion of the atomic pair with $\mu$ the reduced mass and $\ell^2$ the angular momentum. $\hat{H}_{el}$ is the electronic Hamiltonian and $\hat{H}_{SO}$ the spin-orbit coupling. The term $\frac{1}{2}\mu\omega^2_{rel}R^2$ is the harmonic potential for the relative motion in the trap with the frequency $\omega_{rel}=\sqrt{\frac{m_{Sr}\omega_{Rb}^2+m_{Rb}\omega_{Sr}^2}{m_{Rb}+m_{Sr}}}$. Finally, the last term is a dynamical term coupling the two motions, $\Delta\omega =\sqrt{\omega_{\textrm{CM}}^2-\omega_{rel}^2}$ being related to their frequency difference.

The  neglect of the coupling term turns out to be a crucial issue in order to avoid an otherwise complicated theoretical description of the atomic pair. Fortunately, Saenz $\it{et}$ $\it{al}$.\cite{phot_harm_trap_1,pair_atom_harm_trap_9,pair_atom_harm_trap_10} give some regimes where the dynamical term is small and the separation remains a good approximation. This is precisely the case when the ratio $\frac{a}{a_{\omega}}$ is small, where $a_{\omega}=\frac{1}{\sqrt{\mu\omega_{rel}}}$ is the characteristic length of the relative motion in the trap and $a$ the scattering length. For $\omega_{rel}= 2\pi\times 400$~kHz, used in our calculation, $a_{\omega}=484$ $a_0$ is larger than the scattering length, $a=90.9$ $a_0$. The separation of the two motions could then be taken as a good approximation. Further study taking into account this term would be interesting, but in the following, we assume the validity of the separation. 

\subsection{Methodology for the resolution of the relative motion}
After separation, the problem of the relative motion must be solved, with the following Hamiltonian : 
 \begin{equation}
\hat{H}_{trap}^{rel}=-\frac{\hbar^2}{2\mu}\nabla_{R}^2+\hat{H}_{rot}+\hat{H}_{el}+ \hat{H}_{SO}+\frac{1}{2}\mu\omega^2_{rel}R^2.
\end{equation}
\begin{figure}
	\centering
	\includegraphics[width=8cm]{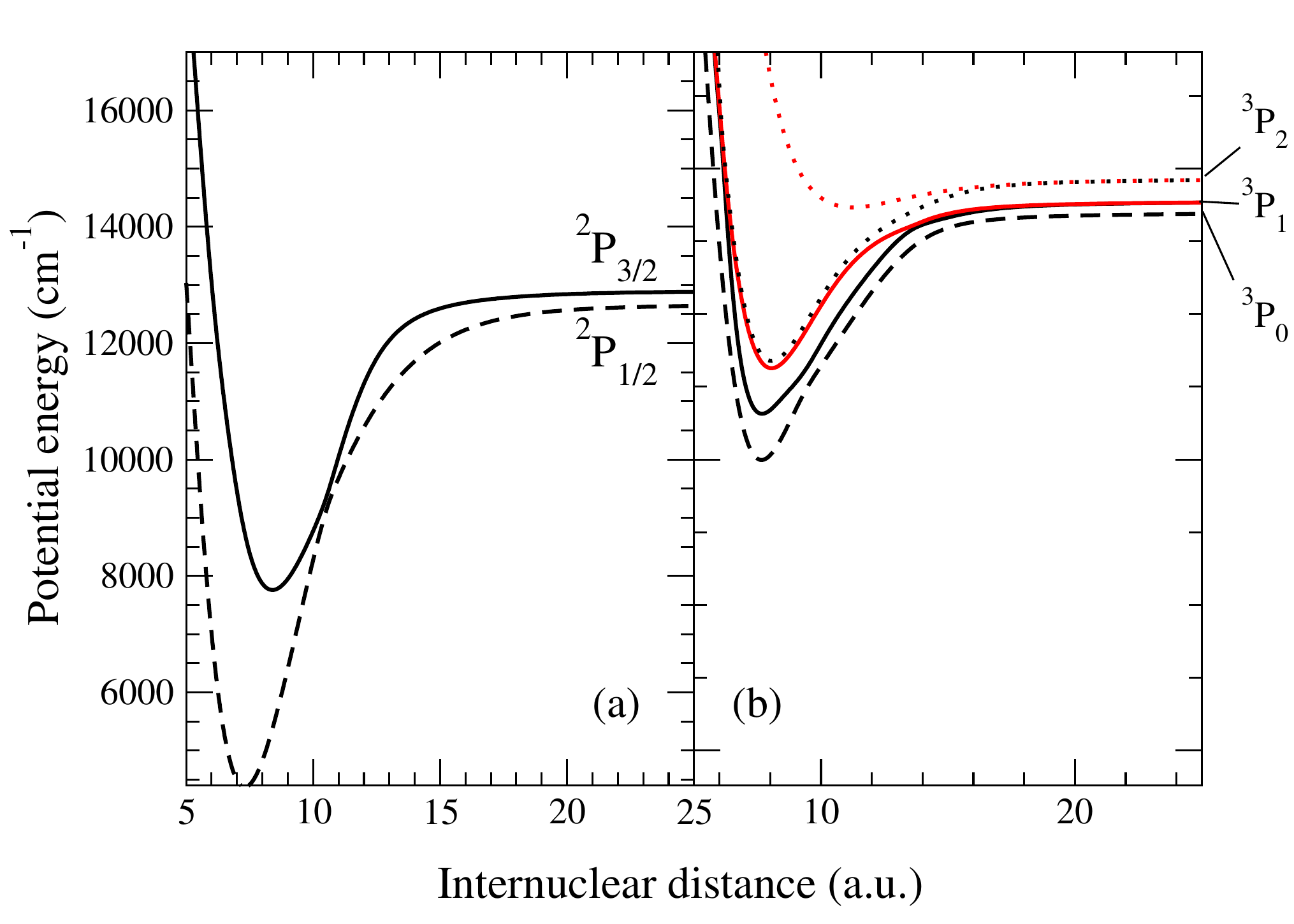}
	\caption{Potential energy curves for the excited electronic states of RbSr with inclusion of spin-orbit couplings. Panel (a) shows the electronic states correlating to the Rb (5p$^2$P$_{1/2,3/2}$)+Sr (5s$^2$ $^1$S) asymptotes, while panel (b) shows the electronic states correlating to the Rb (5s$^2$S)+Sr (5s5p $^3$P$_{0,1,2}$) asymptotes. In panel (a), the dashed line correlates to the Rb (5p$^2$P$_{1/2}$)+Sr (5s$^2$ $^1$S) asymptote and the solid one to the Rb (5p$^2$P$_{3/2}$)+Sr (5s$^2$ $^1$S) asymptote. In panel (b), the dashed line correlates to the Rb (5s$^2$S)+Sr (5s5p $^3$P$_{0}$) asymptote, the solid lines to the Rb (5s$^2$S)+Sr (5s5p $^3$P$_{1}$) asymptote and the dotted lines to the Rb (5s$^2$S)+Sr (5s5p $^3$P$_{2}$) asymptote.}
	\label{fig:PEC}
\end{figure}
The electronic structure calculation has been conducted in detail in our previous papers \cite{RbSr_2,devolder2018}. A full configuration interaction (FCI) method  with effective core potential (ECP) and core polarization potential (CPP) was used.  The inclusion of the spin-orbit coupling is  also explained in our papers \cite{RbSr_2,devolder2018}. The relevant excited electronic states for our study are shown in Figure \ref{fig:PEC}. They correlate to two groups of asymptotes : Rb (5p$^2$P$_{1/2,3/2}$)+Sr (5s$^2$ $^1$S) and Rb (5s$^2$S)+Sr (5s5p $^3$P$_{0,1,2}$). The resolution of the nuclear relative motion is done with the Mapped Fourier Grid Hamiltonian method (MFGH) \cite{MFGH_1,MFGH_2,MFGH_3,MFGH_4,MFGH_5,MFGH_6,MFGH_7,MFGH_8,MFGH_9,MFGH_10}, using a grid extending from $R_{min}=5 a_0$ to $R_{max}=5000 a_0$, covering the spatial extension of the lowest trap level. This grid is the same for the ground and excited states in order to calculate the integral of the transition dipole moment between the bound rovibrational levels of electronic states.

\section{Laser control involving two electronic states}
\begin{figure}
	\centering
	\includegraphics[width=8cm]{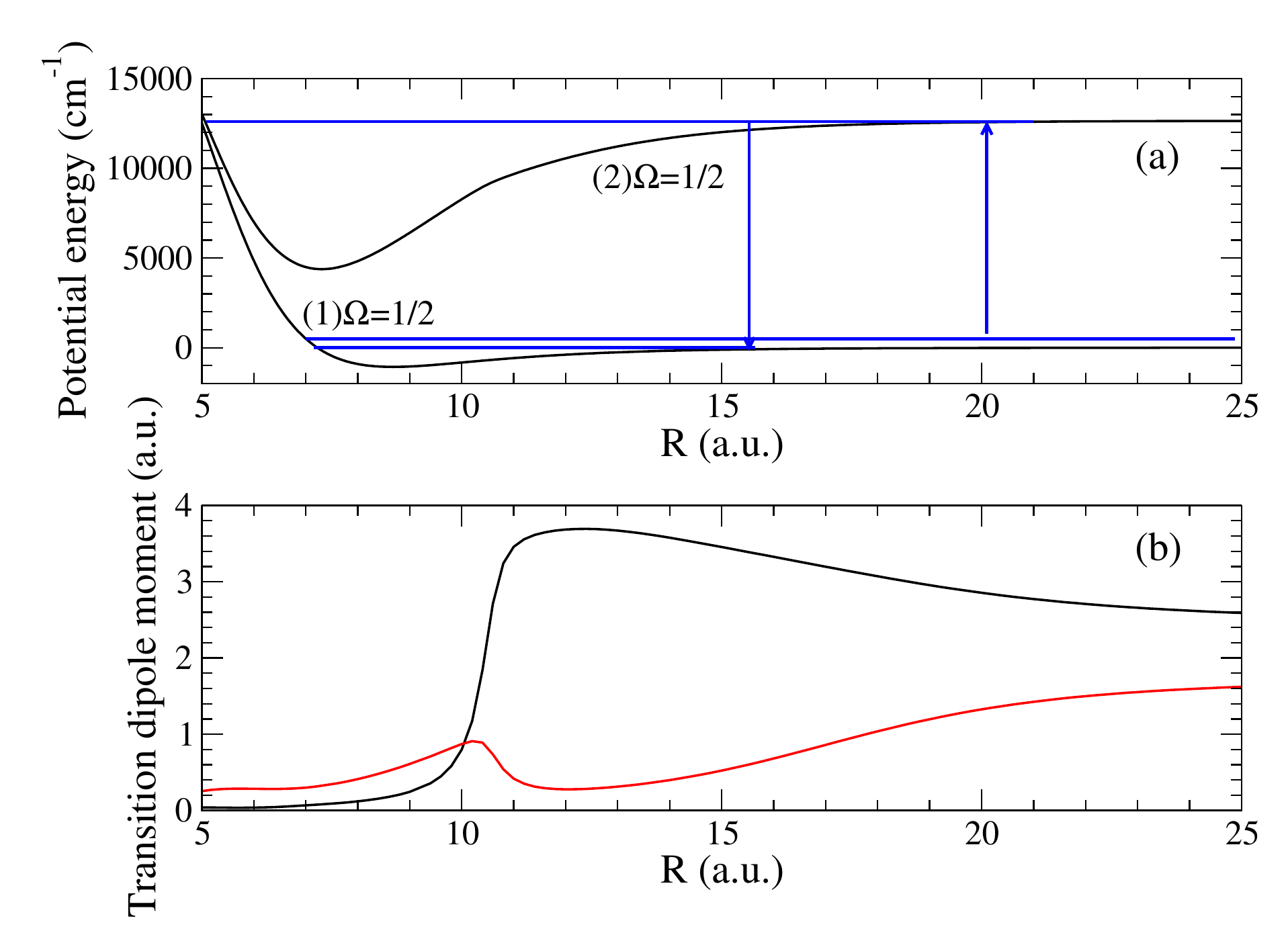}
	\caption{(a) Potential energy curves involved in the first STIRAP transfer between the first trap state and a high rovibrational level of the electronic ground state via a rovibrational level of an excited electronic state ((2)$\Omega=1/2$). (b) Transition dipole moment (TDM) components between the electronic ground excited states. The parallel component $\bra{X^2\Sigma}\mu_{\parallel}\ket{(2)^2\Sigma}$ is plotted in black while the perpendicular one $\bra{X^2\Sigma}\mu_{\perp}\ket{(1)^2\Pi}$ is plotted in red.  }
	\label{fig:pot_1_STIRAP}
\end{figure}

\begin{figure}
	\centering
	\includegraphics[width=8cm]{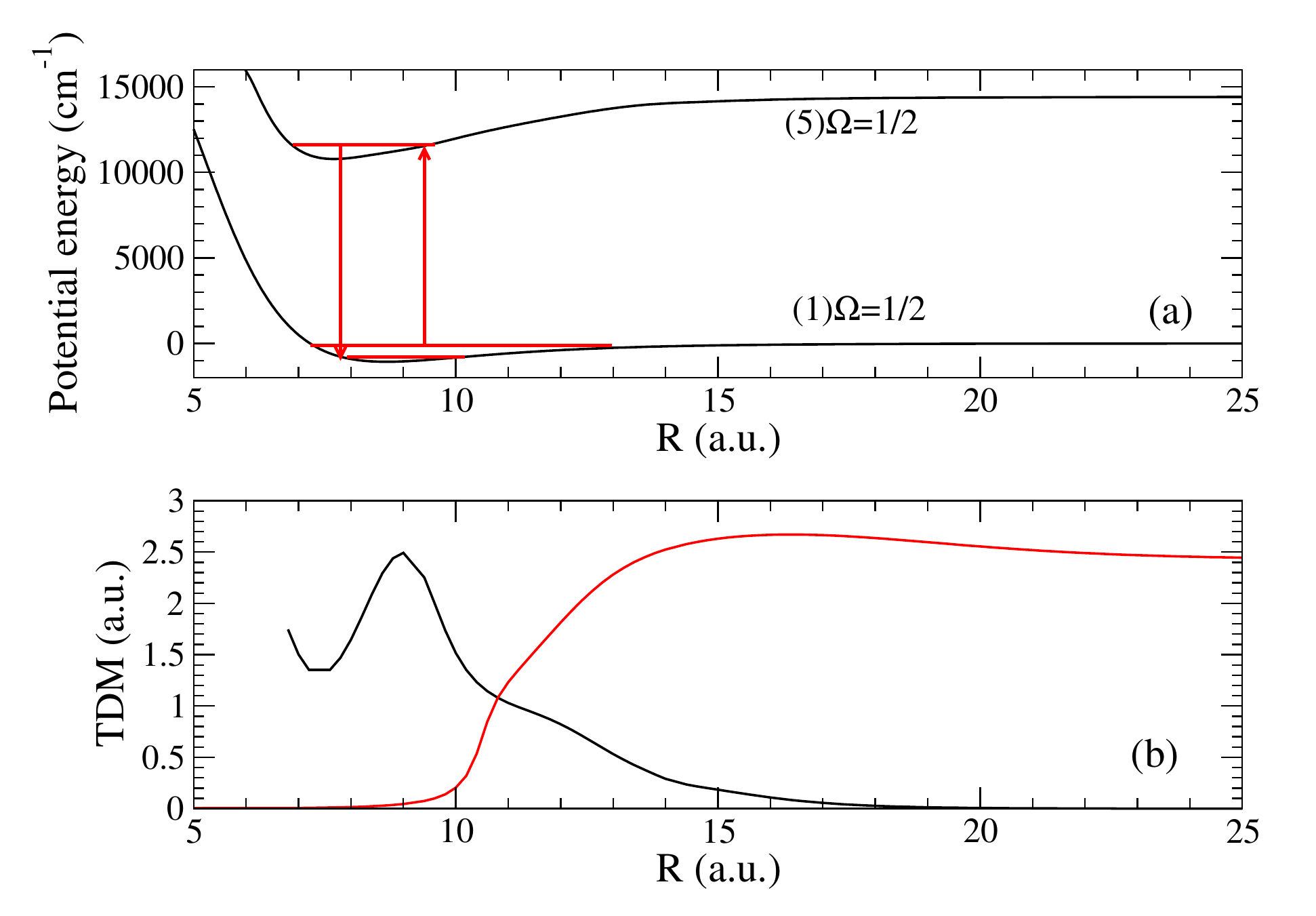}
	\caption{(a) Potential energy curves involved in the second STIRAP transfer between a high rovibrational level of the electronic ground state and the rovibrational ground state via a rovibrational level of an excited electronic state ((5)$\Omega=1/2$). (b) Transition dipole moment (TDM) components between the electronic ground and the excited states. The parallel component $\bra{X^2\Sigma}\mu_{\parallel}\ket{(3)^2\Sigma}$ is plotted in black, while the perpendicular one $\bra{X^2\Sigma}\mu_{\perp}\ket{(2)^2\Pi}$ is plotted in red.}
	\label{fig:pot_2_STIRAP}
\end{figure}
The first strategy studied in this paper for the formation of ultracold RbSr molecules is derivatives of STIRAP, with 5 levels. In our previous paper \cite{devolder2018}, we identified two STIRAP process, one for the chemical bond formation (see Figure \ref{fig:pot_1_STIRAP}) and one for the vibrational stabilization (see Figure \ref{fig:pot_2_STIRAP}). After the first step, the molecules are created in a loosely-bound level, as in magnetoassociation. The goal of the second step is the transfer of the weakly-bound molecules to their rovibrational ground state. The complete story line strategy is sketched in Figure \ref{fig:strategy} (a). 

It is to be noted that the passage through the loosely-bound state during the two-consecutive STIRAP (tc-STIRAP) could induce some problems. More precisely, the loosely bound molecules can be dissociated by different processes like photon scattering or molecule-molecule scattering. After dissociation, the atoms are usually lost from the trap. Fortunately, the risk of the molecule-molecule scattering is already reduced by the Mott-insulator phase, the tunneling of ultracold molecules between sites being negligible. Furthermore, for limiting the risk of losses, we have explored two methods that reduce the population in the weakly-bound level during the process, namely : Alternating-STIRAP (A-STIRAP) and Straddle-STIRAP (S-STIRAP)\cite{Tannor97,Tannor99,Vitanov2017}. In the present case, they correspond to transfers between five levels via four pulses. A S-STIRAP strategy has already been used in the formation of ultra-cold Cs$_2$ dimers \cite{Danzl2010}. The five levels are those of the tc-STIRAP : the first trap level ($N=0,J=1/2$) of the electronic ground state, the vibrational level ($v'=199, J'=3/2$) of the electronic state (2) $|\Omega_{exc}|=1/2$, the vibrational level ($v=-3,J=1/2$) of the electronic ground state, the vibrational level ($v'=15,J'=3/2$) of the state (5) $|\Omega_{exc}|=1/2$ and finally the rovibrational ground state ($v=0, J=1/2$) (negative quantum numbers are counted from the dissociation limit). They are respectively written : $\ket{\Phi_{N=0,J=1/2}^g}$, $\ket{\Phi_{v'=199,J'=3/2}^{(2)|\Omega_{exc}|=1/2}}$, $\ket{\Phi_{v=-3,J=1/2}^g}$, $\ket{\Phi_{v'=15,J'=3/2}^{(5)|\Omega_{exc}|=1/2}}$ and $\ket{\Phi_{v=0,J=1/2}^g}$. Like in the usual three-state STIRAP, the methods are still based on a dark state but of a system with five levels in this case. Within the rotating wave approximation, it is given by :
\begin{widetext}
	\begin{equation}
	\ket{0}=\frac{1}{\mathcal{N}} \left[\Omega_{D_1}(t)\Omega_{D_2}(t)\ket{\Phi_{N=0,J=1/2}^g} -\Omega_{P_1}(t)\Omega_{D_2}(t) \ket{\Phi_{v=-3,J=1/2}^g} +\Omega_{P_1}(t)\Omega_{P_2}(t) \ket{\Phi_{v=0,J=1/2}^g}\right]
	\label{eq:dark_state_A_stirap}
	\end{equation}
\end{widetext}
$\Omega_{P(D)_{1/2}}$ are the Rabi frequencies for the first/second pump (dump) transitions. They are calculated from the transition dipole matrix elements (TDM). After use of angular momentum algebra, we obtain the following expression for the TDM with linear polarization $\hat{\epsilon}_0$ : 
\begin{widetext}
	\begin{equation}
	\begin{split}
	|\bra{\Phi^{g}_{v,J=1/2}}\vec{\mu}.\hat{\epsilon}_0\ket{\Phi_{v',J'=3/2}^{(n)|\Omega_{exc}|=1/2}}|^2=&\frac{2}{9}|\bra{\phi^{X^2\Sigma}_{v,J=1/2}}\bra{X^2\Sigma}\mu_{\parallel}\ket{(n_\Sigma)^2\Sigma}\ket{\phi^{(n_\Sigma)^2\Sigma}_{v',J=3/2}}|^2\\ +&\frac{1}{9}|\bra{\phi^{X^2\Sigma}_{v,J=1/2}}\bra{X^2\Sigma}\mu_{\perp}\ket{(n_\Pi)^2\Pi}\ket{\phi^{(n_\Pi)^2\Pi}_{v',J'=3/2}}|^2.
	\label{eq:TDM_STIRAP}
	\end{split}
	\end{equation}
\end{widetext}
The levels are expressed in the Hund case (a) basis : $\ket{\Phi^{g}_{v,J=1/2}}=\ket{\phi^{X^2\Sigma}_{v,J=1/2}}\ket{X^2\Sigma}$ for the electronic ground state and $\ket{\Phi_{v',J_{exc}=3/2}^{(n)|\Omega_{exc}|=1/2}}=\ket{\phi_{v',J'=3/2}^{(n_\Sigma)^2\Sigma}}\ket{(n_\Sigma)^2\Sigma}+\ket{\phi_{v',J'=3/2}^{^2\Pi}}\ket{(n_\Pi)^2\Pi}$ for the electronic excited states. The quantity $\ket{\phi_{v',J'=3/2}^{(n_\Sigma)^2\Sigma}}$ is the component of this level on the corresponding $^2\Sigma$ state and the same for the $^2\Pi$  state. For the (2) $|\Omega_{exc}|=1/2$ electronic state, $n_\Sigma=2$ and $n_\Pi=1$ while for the (5) $|\Omega_{exc}|=1/2$ electronic state, $n_\Sigma=3$ and $n_\Pi=2$. $\mu_{\parallel}$ and $\mu_{\perp}$ are the parallel and perpendicular components of the dipole moment. The former induces a coupling with the $^2\Sigma$  component of the electronic excited states while the latter leads to a coupling with their $^2\Pi$ component. The matrix elements $\bra{X^2\Sigma}\mu_{\parallel}\ket{(n_\Sigma)^2\Sigma}$ and $\bra{X^2\Sigma}\mu_{\perp}\ket{(n_\Pi)^2\Pi}$ for the involved excited electronic states are represented in Figures \ref{fig:pot_1_STIRAP} (b) and \ref{fig:pot_2_STIRAP}(b). Note that in addition to what has been done in our previous paper \cite{devolder2018}, here we include the rotational contributions. The values (in a.u.) of the TDM for the 4 transitions are the following : $|\mu_{P_1}|^2=5.2 \times 10^{-5}$, $|\mu_{D_1}|^2=1.3 \times 10^{-4} $, $|\mu_{P_2}|^2=5.5 \times 10^{-5}$ and $|\mu_{D_2}|^2=9.8 \times 10^{-5}$.

\begin{figure}
	\centering
	\includegraphics[width=8cm]{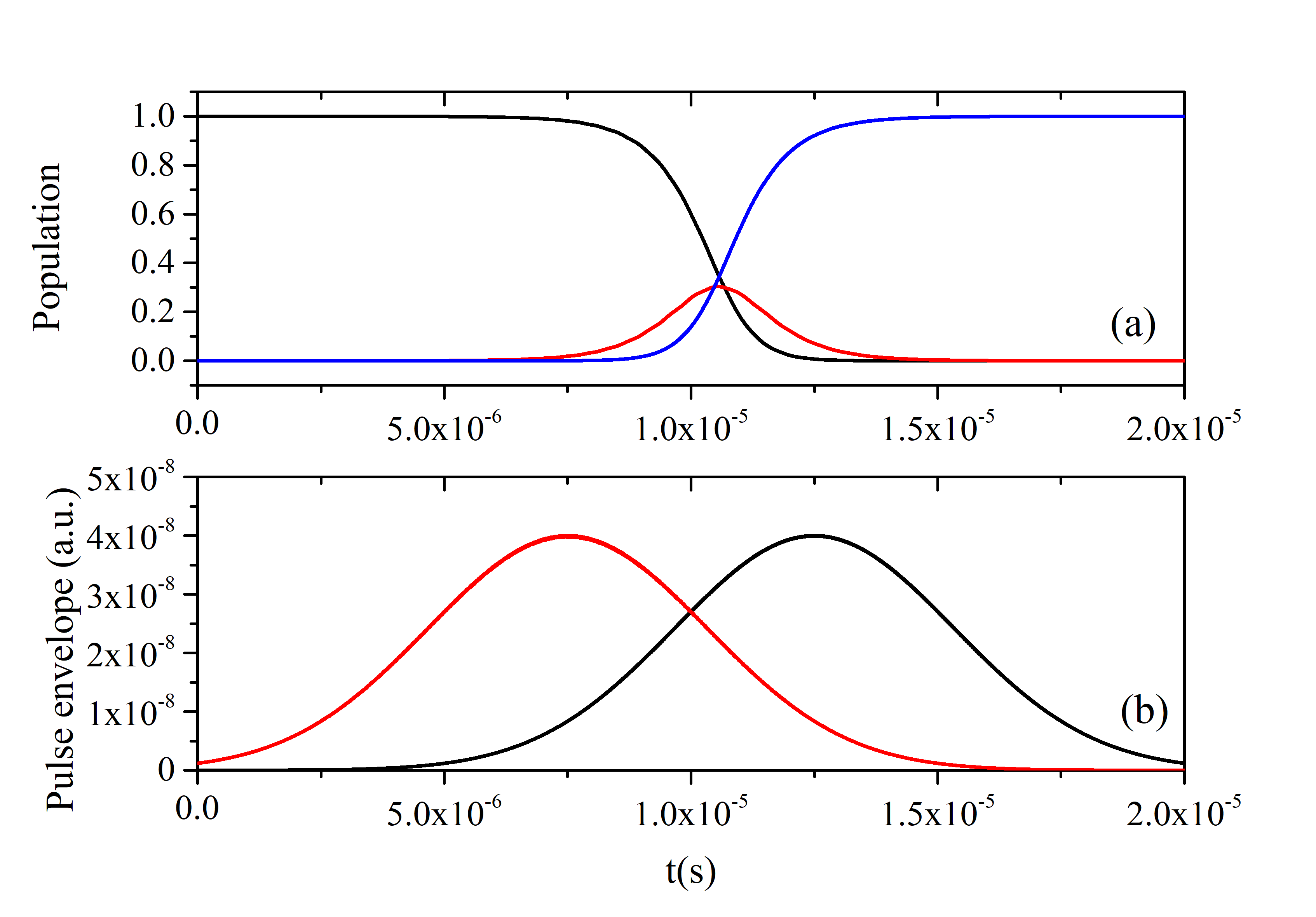}
	\caption{(a) Simulation of population transfer from the first trap level (black) to the rovibrational ground state (blue) via  the loosely bound rovibrational level $v=-5$ (red). (b) Envelope of the pump (black) and dump (red) fields of the A-STIRAP process.}
	\label{fig:A-STIRAP}
\end{figure}

Like in the three-state STIRAP, the principle of transfer is populating the dark state during the whole process and changing the composition of this dark state. The A-STIRAP consists in applying the two dump pulses (second and fourth transitions) prior to the two pump pulses (first and third transitions). We have simulated the dynamics of the A-STIRAP process. In interaction representation, the total wavefunction is expressed on a basis of rovibrational  and trap levels of the electronic states $\ket{\Phi_{v_k,J_k}^{el_k}}$ :
\begin{equation}
\ket{\Psi(t)}=\sum_k \tilde{C}_k (t) e^{-\frac{i}{\hbar}E_kt}\ket{\Phi_{v_k,J_k}^{el_k}}.
\end{equation}
The time-dependent coefficients $\tilde{C}_k (t)$ are obtained by resolution of the coupled equations :
\begin{equation}
i\hbar \frac{d\tilde{C}_{k}}{dt}=-\sum_{j}\mu_{kj} \mathcal{E}(t) e^{\frac{i}{\hbar}(E_{k}-E_{j})t}\tilde{C}_{j}(t)
\label{eq:STIRAP_Interaction_picture}
\end{equation}
where $\mu_{kj}=\bra{\Phi_{v_k,J_k}^{el_k}}\mu. \hat{\epsilon}_0\ket{\Phi_{v_j,J_j}^{el_j}}$. \\
The propagation of these coupled equations is made with a Runge-Kutta 4 (RK4) algorithm with a time step of 24 fs. The field comprises four pulses, with the following form :
\begin{equation}
\begin{split}
\mathcal{E}(t)=&\mathcal{E}_{P_1} \exp\left(-\frac{(t-t_{c}^{P_1})^2}{\tau^2_{P_1}}\right)\cos(\omega_{P_1}t)\\+&\mathcal{E}_{D_1} \ \exp\left(-\frac{(t-t_{c}^{D_1})^2}{\tau^2_{D_1}}\right)\cos(\omega_{D_1}t)\\
+&\mathcal{E}_{P_2} \exp\left(-\frac{(t-t_{c}^{P_2})^2}{\tau^2_{P_2}}\right)\cos(\omega_{P_2}t)\\+&\mathcal{E}_{D_2} \ \exp\left(-\frac{(t-t_{c}^{D_2})^2}{\tau^2_{D_2}}\right)\cos(\omega_{D_2}t),
\end{split}
\label{eq:champ_A-STIRAP}
\end{equation}
where $\mathcal{E}_{(P,D)_1}$ are the maximum field amplitudes, $t_{c}^{(P,D)_1}$ are the Gaussian pulses central times, $\tau_{(P,D)_1}$ are the temporal widths and $\omega_{(P,D)_1}$ are the frequencies of the transition. Indices 1 and 2 correspond to the transitions involved in the first and second STIRAP respectively. 
The simulation of the population dynamics is conducted with a basis of 33 levels made of the five levels of the A-STIRAP transfer, to which are added 20 excited trap states $(N=0-19,J=1/2)$, the rovibrational levels $(v'=198,J=3/2)$ and $(v'=200,J=3/2)$ of (2)$|\Omega_{exc}|=1/2$, the vibrational levels $(v'=14,J=3/2)$ and $(v'=16,J=3/2)$ of (5)$|\Omega_{exc}|=1/2$, and the vibrational levels $(v=-4,J=1/2)$ and $(v=-2,J=1/2)$ of the electronic ground state. A complete transfer is found for a pulse of 20 $\mu$s (see Figure \ref{fig:A-STIRAP}). The amplitudes of the pulses are $4.0 \times 10^{-8}$ a.u. (which amounts to intensities of 306 W/cm$^2$). Such intensities  are potentially reachable in current ultracold experiments. The other field parameters take the following values : $t_{c}^{P1}=t_{c}^{P2}=12.5 \mu s$, $t_{c}^{D1}=t_{c}^{D1}=7.5 \mu s$ and $\tau_{P1}=\tau_{P2}=\tau_{D1}=\tau_{D2}=4.0 \mu s$. 

\begin{figure}
	\centering
	\includegraphics[width=8cm]{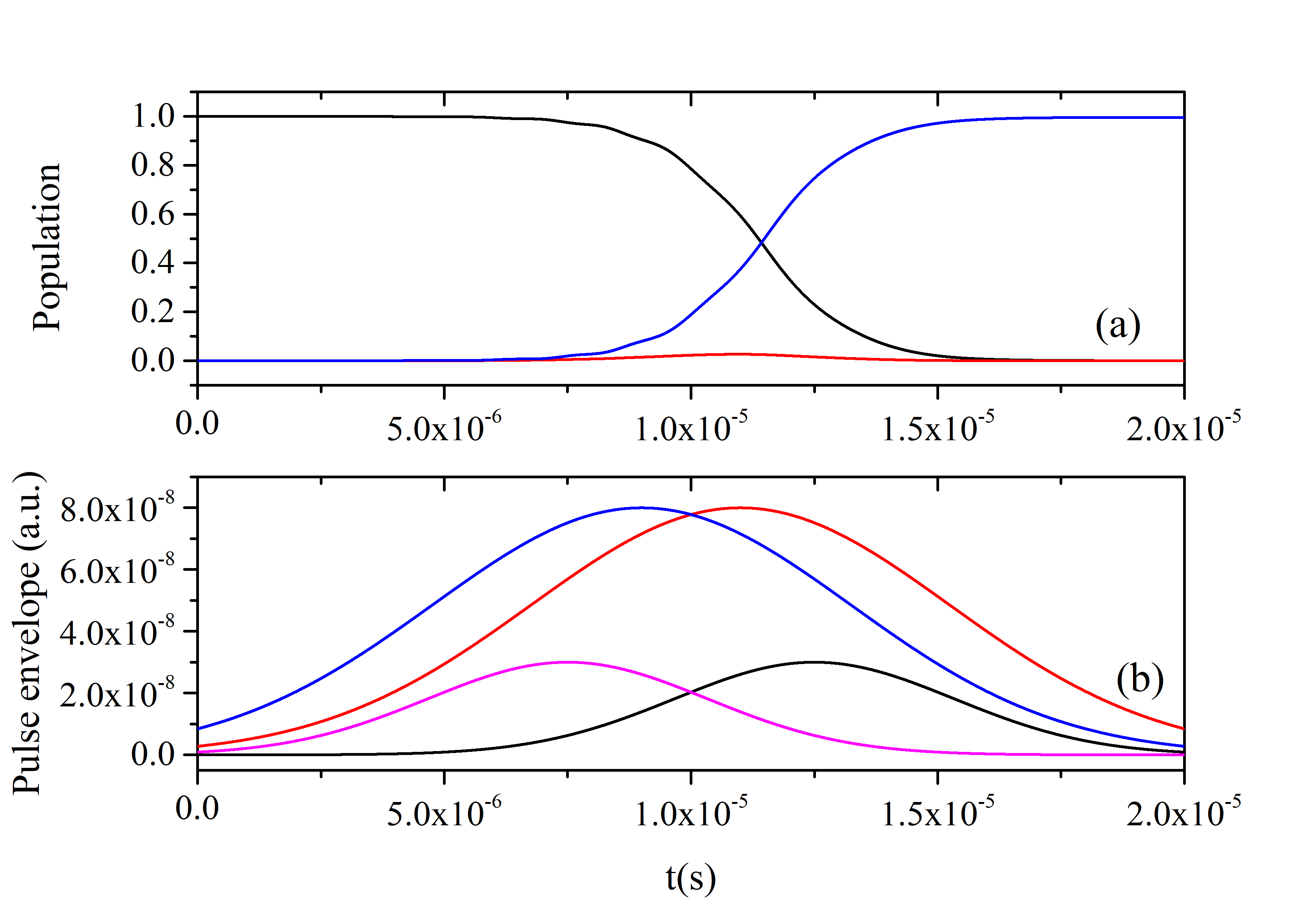}
	\caption{(a) Simulation of population transfer from the first trap level (black) to the rovibrational ground state (blue) via  the loosely bound rovibrational level $v=-5$ (red). (b) Envelope of the first pump (black), first dump (blue), second pump (red) and second dump (magenta) fields of the S-STIRAP process.}
	\label{fig:S-STIRAP}
\end{figure}

Unfortunately, the A-STIRAP does not solve completely the issue of the occupation of loosely-bound level ($v$=-3). This is expected since this level has a component in the dark state \ref{eq:dark_state_A_stirap}. The maximum population of this level during the dynamics is 30 \% which is still an improvement with respect to the tc-STIRAP where 100 \% occupies this level between the two STIRAPs. 
With the S-STIRAP, we can go further since we succeed in minimizing the transitory population in the excited state. In this case, the second (first dump) and third (second pump) pulses have higher amplitudes and completely overlap the first (first pump) and the fourth (second dump) pulses (see Figure \ref{fig:S-STIRAP} (b)). Again, we obtain a efficient transfer for a pulse of 20 $\mu$s. The parameters of the pulses are the following : $\mathcal{E}_{P1}=5.0 \times 10^{-8}$ a.u., $t_{c}^{P1}=12.5 \mu s$, $\tau_{P1}=5.7 \mu s$; $\mathcal{E}_{D1}=8.0 \times 10^{-8}$ a.u., $t_{c}^{D1}=11 \mu s$, $\tau_{D1}=8.5 \mu s$; $\mathcal{E}_{P2}=8.0 \times 10^{-8}$ a.u., $t_{c}^{P2}=9 \mu s$, $\tau_{P2}=8.5 \mu s$; et $\mathcal{E}_{D2}=5.0 \times 10^{-8}$ a.u., $t_{c}^{D2}=7.5 \mu s$, $\tau_{D2}=5.7 \mu s$. The peak intensities are 1220 W/cm$^2$ which are  more challenging to achieve on such pulse duration.  However, the S-STIRAP allows one to minimize the population in the weakly-bound level ($v$=-3). The maximum population in this level is 2.7 \% and can be even more decreased if we still increase the amplitudes $\mathcal{E}_{D1}$ and $\mathcal{E}_{P2}$. In conclusion, the S-STIRAP seems the best solution if the lifetime of the loosely-bound level is very short and is a strong constraint on the transfer. The cost would be the use of strong field intensities. In other situations, the A-STIRAP seems the most appropriate method since it requires lower laser intensities. 
 
\section{Laser control involving a single electronic (ground) state}
\begin{figure}
	\centering
	\includegraphics[width=8cm]{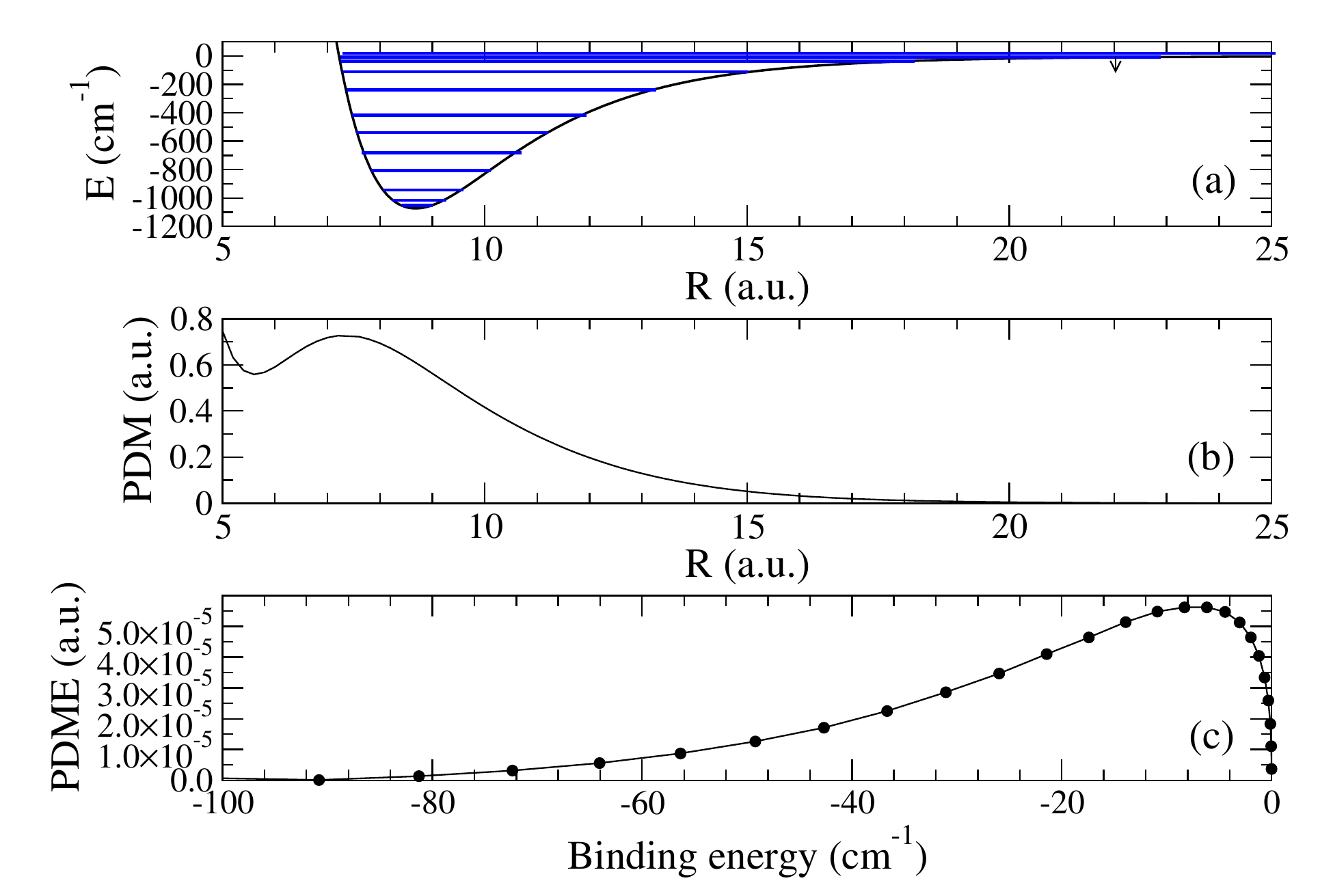}
	\caption{(a) Potential energy curve of the electronic ground state. Vibrational states, which are involved in the transfer to the  rovibrational ground state, are illustrated by blue solid lines. (b) Permanent dipole moment (PDM) $\bra{X^2\Sigma}\mu_{\parallel}\ket{X^2\Sigma}$ of the electronic ground state. (c) PDM matrix element (PDME) between the first trap level $N=0$ and the vibrational levels of the electronic ground state (without the rotation) } 
	\label{fig:mom_dip_perm}
\end{figure}

Another solution for avoiding the spontaneous emission is restricting the dynamics to the electronic ground state. The formation of heteronuclear molecules is possible via this strategy thanks to the presence of a permanent dipole moment (represented in  Figure \ref{fig:mom_dip_perm} (b) for the electronic ground state of RbSr). Unfortunately, the matrix element between the first trap level and the rovibrational ground state is too small ($10^{-20}$ a.u.) for inducing a direct transfer. Like for the STIRAP, the strategy must be divided in two parts : the chemical bond formation and the vibrational stabilization. 
\subsection{Chemical bond formation}
First, we must identify the final rovibrational level. The criterion for that purpose is simply the vibrational level with the highest PDM matrix element with the first trap level, which actually is for the vibrational level $v$=56 \cite{devolder2019} (see Figure \ref{fig:mom_dip_perm}(c)). Concerning the rotation, the highest factors are obtained for a transition $J=1/2 \rightarrow 3/2$. For a linear polarisation, the angular factor is $\sqrt{2}/3$. Two types of pulses are considered for the transfer: a chirped pulse and a $\pi$-pulse.
\subsubsection{Chirped pulse}
The principle of the transfer is the adiabatic passage between the two levels via a chirped pulse \cite{Guerin2011,Malinovsky2001}. In the dressed representation, the two levels cross each other while there is an avoiding crossing between the two adiabatic levels. When the frequency is slowly swept across the resonance in order that the system always stays in the same adiabatic level, the system is transferred from the first trap level to the target rovibrational level. 

We note that this method of formation follows the same principle than the magnetoassociation  based on a magnetic Feshbach resonance. In a similar way, the laser coupling between the trapping level and the vibrational level can be related to an optical Feshbach resonance, that we recently identified as a Laser Assisted Self Induced Feshbach Resonance (LASIFR) \cite{devolder2019}. Besides the molecule formation, by changing the frequency, the scattering length can be controlled. Conceptually, the LASIFR concept leads to the idea that the field-coupling of continuum levels with a bound level of the same electronic potential can be related to a Feshbach resonance in the field representation.

As we can expect, the formation of the molecule can be described by the Landau-Zener (LZ) model. 
\begin{equation}
\begin{cases}
p_{trap}=exp(-\pi\frac{\Omega_{LASIFR,max}^2}{2|\alpha|}) \\
p_{v}=1-exp(-\pi\frac{\Omega_{LASIFR,max}^2}{2|\alpha|})
\end{cases}.
\label{LZ_model}
\end{equation}
where $\Omega_{LASIFR,max}$ is the maximum Rabi frequency at the avoided crossing.  $p_v$ is the final probability of the vibrational level while $p_{trap}$ is the final probability of the trap level.  $\alpha$ is the linear chirp rate and is defined by:
\begin{equation}
\alpha=\frac{{{\omega }_{f}}-{{\omega }_{0}}}{{{t}_{\max }}}
\label{chirp_rate}
\end{equation}
  where $\omega_{0,f}$ are the initial and final frequency, respectively and $t_{max}$ is the duration of the pulse. \\
However, the LZ model has limits and before going further, we discuss these limitations in this paragraph. The main  origin is the two-state assumption. Only the trap level and the vibrational level are considered in the LZ model. However, the transfer can be perturbed by other levels, and specially by the other trap levels. A first consequence of the inclusion of trap levels is the asymmetry in the sign of the ramp. An efficient transfer can only be obtained for a negative ramp. A positive ramp induces the transfer to the other trap level and hence induces a heating of the ultracold gas. For a negative ramp, there is also a limitation due to the last vibrational level. If the variation of the frequency is greater than the binding energy of the last vibrational level, the population is transferred to this level and not to the target vibrational level. These constraints on the sign and the magnitude of the ramp are also observed for the magnetoassociation \cite{Magnetic_Feshbach_resonance_2}. Furthermore, the presence of the other trap level also induces a Stark shift on the two-levels involved in the adiabatic passage. The value of this Stark-shift can be reduced with a high  trap frequency, like in the Mott insulator phase. The increase of the trap frequency also allows the increase of the coupling between the first trap level and the target vibrational level ($v$=56). 

The numerical implementation of the LZ model (\ref{LZ_model}) is shown in table \ref{tab:lz_pol} for a linear polarization. The adiabaticity transfer is fulfilled when $\Omega_{LASIFR,max}^2 >> \alpha$. The comparison between transfer strategies is made referring to the maximal intensity which is the main experimental constraint. For an intensity similar to that used for the STIRAP (hundred of W/cm$^2$), a pulse duration of 10 ms is necessary. The longer duration of the pulse is due to the smaller values of permanent versus treansition dipole moments.  An interesting advantage with respect to the tc-STIRAP of Section III,  is that the created molecules occupy a more bounded rovibrational level ($v$=56 against $v$=63), less sensitive to losses. 

\begin{table}
	\centering
	\begin{tabular}{|c|c|c|}
		\hline
		$t_{max} (\mu s)$ & $|\alpha|$ ($MHz/\mu s$) & $I_0$ (W/cm$^2$)  \\
		\hline
		100 &1.0 $\times10^{-2}$  & 30000  \\
		100 &5.0 $\times 10^{-3}$ &  15000   \\
		10000&1.0 $\times 10^{-4}$  & 300   \\
		10000&5.0 $\times 10^{-5}$ & 150  \\
		10000&5.0 $\times 10^{-5}$  & 225  \\
		\hline	 
	\end{tabular}
	\caption{Values of the parameters inducing a population of 99 \% in the rovibrational level ($v=56,J=3/2$) via an adiabatic passage from the first trap state ($N=0,J=1/2$). $\alpha$ : is the chirp rate; $I_0\propto \mathcal{E}_0^2$: the peak intensity; $t_{max}$ the duration of the pulse. The laser is linearly polarized.}
	\label{tab:lz_pol}
\end{table}

\subsubsection{$\pi$-pulse}
An alternative to a chirped pulse is a $\pi$-pulse \cite{Thomas1983,Holthaus1994}. The principle of the $\pi$ pulse is that the temporal integral of the Rabi frequency is a multiple of $\pi$. Contrary to the chirped pulse, the frequency is not time-dependent and is fixed at resonance. The field has a Gaussian shape like in Eq.(\ref{eq:champ_A-STIRAP}), with similar parameters specification:
\begin{equation}
\mathcal{E}(t)=\mathcal{E}_0 \ exp\left(-\frac{(t-t_{c})^2}{\tau^2}\right)\cos\left(\omega_0t\right).
\end{equation}
For such a linearly polarized pulse, we can derive an analytical formula for the conditions of a complete transfer :
\begin{equation}
\mathcal{E}_0 .\tau=\frac{\sqrt{\pi}}{\bra{\phi_{v_{res}=56,J=3/2}^g}\bra{X^2\Sigma}\mu_{\parallel}\ket{X^2\Sigma}\ket{\phi_{N=0,J=1/2}^g}}.
\label{cond:pi_pulse_2}
\end{equation}
In Table \ref{tab:pi_pol_lin}, we give the couple ($\mathcal{E}_0$,$\tau$) which fulfill the condition (\ref{cond:pi_pulse_2}). The required intensities are lower than for the chirped pulses. As the intensity is a major limitation for Terahertz sources, it is a clear advantage for $\pi$-pulses with respect to the chirped ones. On the other hand, the $\pi$-pulses are more sensitive to noises affecting laser parameters. The adiabatic passage is a more robust method. In conclusion, the $\pi$-pulse is now the more feasible and the more realistic method but we can expect that with the development of Terahertz sources, chirped pulses could become the privileged method. 

\begin{figure}
	\centering
	\includegraphics[width=8cm]{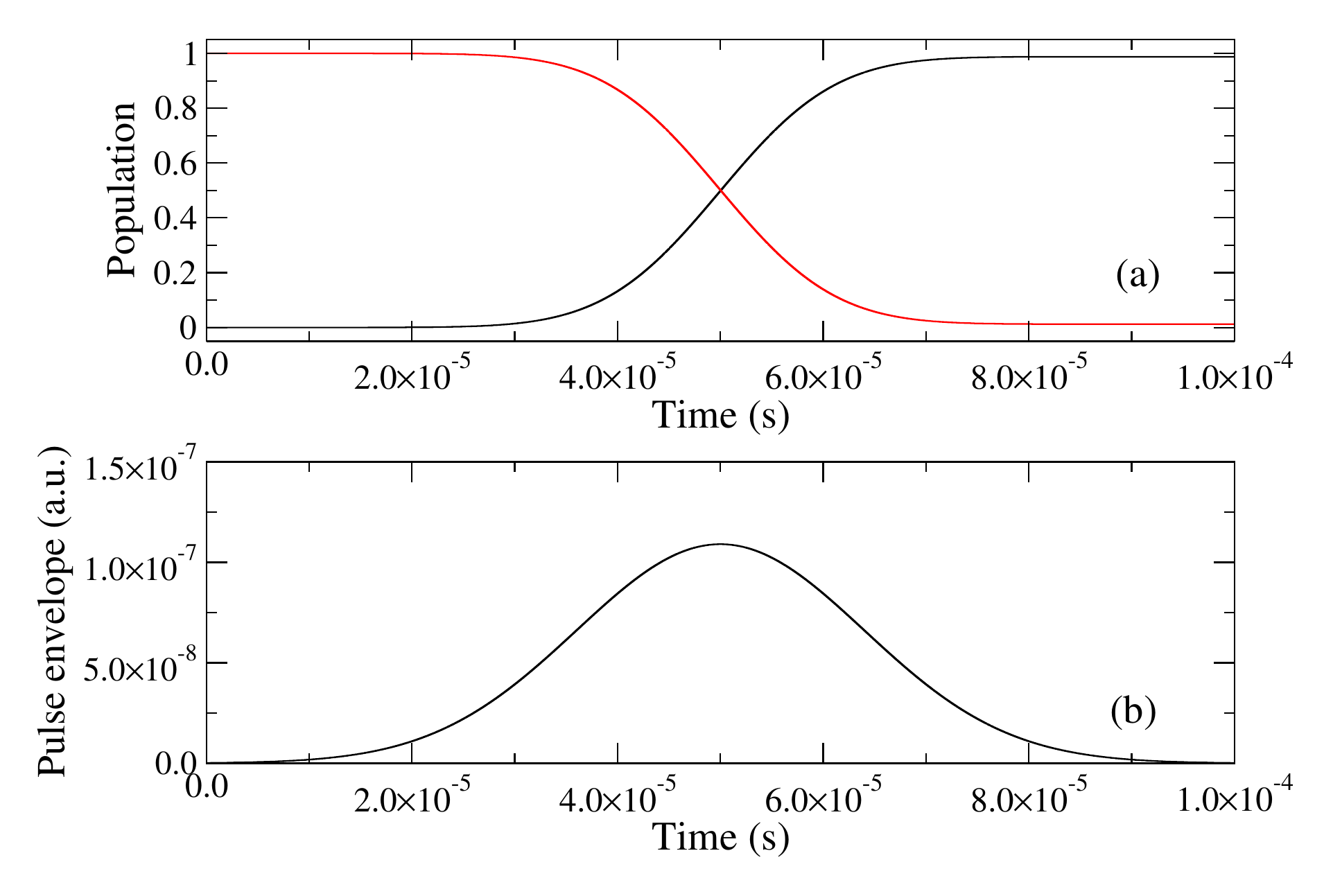}
	\caption{(a) Evolution of the populations in the first trap state (black) and in the target rovibrational level (red) induced by a $\pi$-pulse whose the envelope is shown in (b){\tiny }}
	\label{fig_pi_pulse}
\end{figure}

Finally, we have simulated the $\pi$-pulses with the parameters of Table \ref{tab:pi_pol_lin}. We follow the same methodology than for the simulation of the STIRAP process. We used an extended basis of 20 trap states ($N=0-19,J=1/2$) and 4 vibrational levels ($v=55, 56, 57$ and $65,J=3/2$). The simulation for $\tau=19.8\mu$s and $\mathcal{E}_0=1.09 \times 10^{-7}$ a.u. is illustrated in Figure \ref{fig_pi_pulse}. The final population in the target vibrational level is 98.8 \%, illustrating the efficiency of the transfer. The perturbation by the other trap states is limited and does not affect the efficiency of the transfer. 

\begin{table}
	\centering
	\begin{tabular}{|c|c|c|c|}
		\hline
		$t_{max}$ ($\mu$s) & $\tau$ ($\mu$s)  & $\mathcal{E}_0$ (a.u.) & $I_0$ (W/cm$^2$)  \\
		\hline
		10 & 2.0 & 1.09 $\times 10^{-6}$  & 123987 \\
		20 & 3.4 & 5.44 $\times 10^{-7}$  & 30997 \\
		50 & 9.9 & 2.18 $\times 10^{-7}$  & 4959 \\
		100 & 19.8 & 1.09 $\times 10^{-7}$  & 1240 \\
		200 & 39.6 & 5.44 $\times 10^{-8}$  & 310 \\
		500 & 99.0 & 2.18 $\times 10^{-8}$  & 50 \\
		1000 & 198.0 & 1.09 $\times 10^{-8}$  & 12 \\
		10000 & 1980.0 & 1.09 $\times 10^{-9}$  & 0.12 \\
		\hline	 
	\end{tabular}
	\caption{ Values of the parameters of a Gaussian pulse for obtaining a $\pi$ pulse inducing the transition from the trap level ($N=0,J=1/2$) to the rovibrational state ($v=56,J=3/2$). A linear polarization is used.}
	\label{tab:pi_pol_lin}
\end{table}

\subsection{Vibrational stabilization}
\begin{figure}
\centering
\includegraphics[width=8cm]{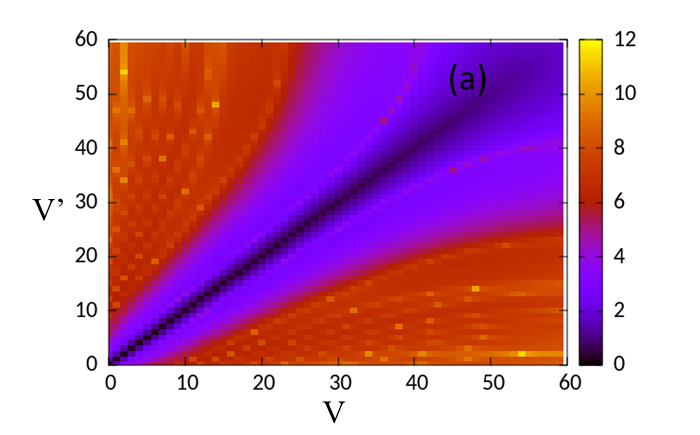}
\includegraphics[width=8cm]{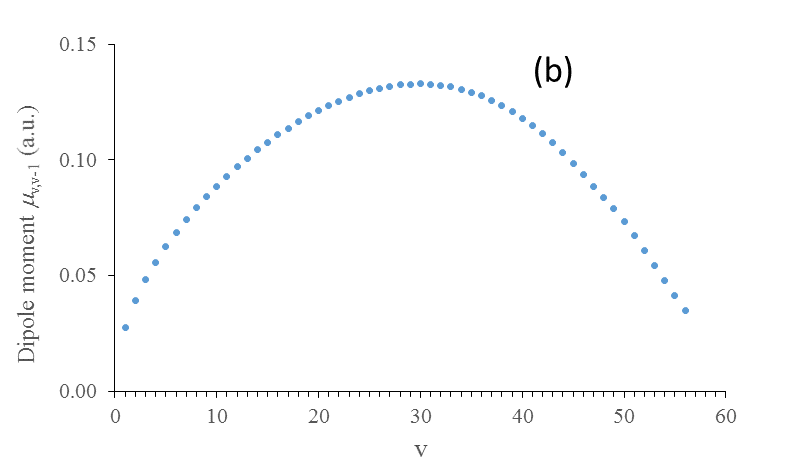}	
\caption{Panel (a) : $-{{\log }_{10}}({{\mu }_{vv'}})$ of the matrix elements of the dipole moment in a.u. for 60 eigenstates for a trap of 400 KHz; panel (b): dipole moment ${{\mu }_{v,v-1}}$ between neighboring states.}
\label{fig_dip}
\end{figure}

In this section, we examine the pros and cons of different strategies in the infra-red domain that would in principle allow the transition from the molecular bound level $v = 56$ prepared in the previous subsection. However, generating shaped pulses in this spectral region remains very challenging nowadays. This simulation aims at analyzing the limits of some alternatives accounting for the complexity of the vibrational chain and wide order of magnitude of the dipolar coupling and by using a reasonable intensity range. In a first attempt, we disregard the Clesh-Gordan coefficients and the rotational degree of freedom. The basis set contains 80 states, i.e. we retain the very dense manifold above the $v = 56$ level. The computation is carried out by the Runge Kutta algorithm in interaction representation without rotating wave approximation. The time step is calibrated from the highest frequency involved in this simulation and fixed to 0.12 ps. 
Figure \ref{fig_dip} displays the matrix elements $\mu _{vv'}$ of the permanent dipole moment.
We compare two strategies. In the first one, in order to minimize the number of pulses we select a chain of about ten states for which the dipolar coupling is smaller than the value among neighboring states but is larger than $10^{-4}$ a.u. We then compare a succession of $\pi$-pulses with a S-STIRAP sequence improved by optimal control (OCT) \cite{Brit2010} according to the Rabitz iterative scheme \cite{Rabitz98} where the S-STIRAP field is taken as the guess field. In the second approach, we exploit all the successive neighboring states by using a chirped pulse. The latter is also use as a guess field for improving the yield by OCT.

\subsubsection{Train of IR $\pi$-pulses}
\begin{figure}
\centering
\includegraphics[width=8cm]{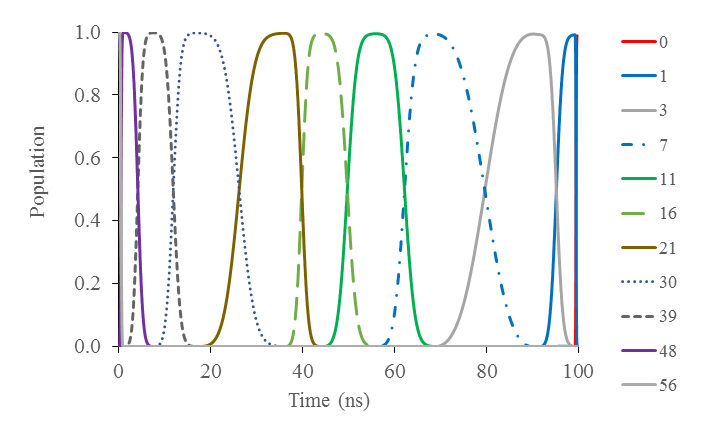}
\includegraphics[width=8cm]{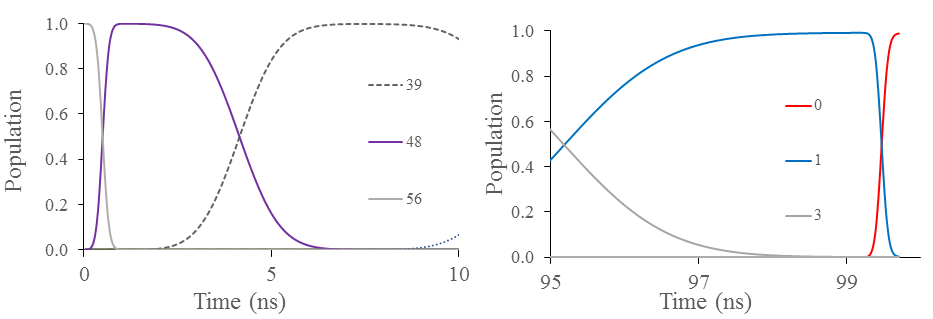}
\caption{Upper panel: Population evolution driven by concatenated $\pi$-pulses described in Table III of the Appendix. Lower panel: Zoom on the two short pulses inducing the first and last transitions.}
\label{fig_poppi}
\end{figure}

The chosen intermediary transitions and the corresponding dipolar couplings are given in Table III of the Appendix. The electric fields are once more written as:
\begin{equation}
{{\mathcal{E}}_{ij}}(t)={{\mathcal{E}}_{0,ij}}\exp (-{{(t-{{t}_{c,ij}})}^{2}}/\tau _{ij}^{2})\cos ({{\omega }_{ij}}t).
\end{equation}
where the indices $ij$ corresponds to the transition $i\rightarrow j$. \\
In a first attempt, the leading amplitude is fixed at $10^{-5}$ a.u. (corresponding to $3.51 \times 10^6$ W/cm$^2$) for each transition and the width ${{\tau }_{ij}}=\sqrt{\pi }/({{\mu }_{ij}}{{\mathcal{E} }_{0,ij}})$ is adjusted to satisfy the $\pi $-pulse condition in a.u.
\begin{equation} 
\int_{0}^{{{t}_{\max ,ij}}}{{{\mu }_{ij}}{{\mathcal{E}}_{ij}}(t)}dt=\pi.
\end{equation}
The final time ${{t}_{\max ,ij}}$ for the simulations is taken as ${{t}_{\max ,ij}}=4\sqrt{\ln (2)}{{\tau }_{ij}}$. As the couplings are varying by two orders of magnitude along the selected chain, the pulse durations strongly differ and the width ${{\tau }_{ij}}$ are given in Table III of the Appendix. Obviously, an alternative should be to fix the pulse duration and let the amplitude vary. Each $\pi$-pulse ensures the selected transition without any perturbation due to the other states of the basis set. We have concatenated all the pulses in order to check the stability and loss of phase relation since every simulation were independent without control of the final phase of the probability amplitude in the final state. The vibrational ground state is reached within $98.9\%$ when the full transition is driven by the concatenated field. The population evolution is displayed in Figure \ref{fig_poppi}. The complete transition requires about 100 ns with the chosen maximum amplitude. The first and last transition have a short duration due to their high dipole moment.  

\subsubsection{S-STIRAP and OCT}
\begin{figure}
\centering
\includegraphics[width=8cm]{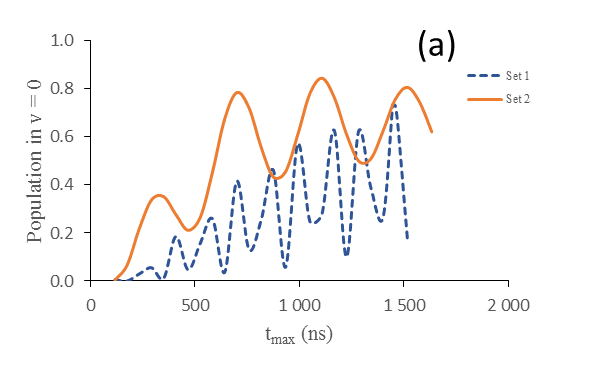}
\includegraphics[width=8cm]{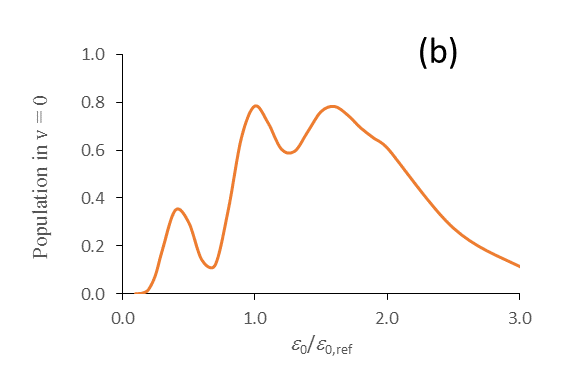}
\caption{Panel (a) : Population in the ground $v$ = 0 state as a function of the total sequence duration $t_{max}$ for two parameter sets. Dashed line: set 1 given in Table \ref{table:S-STIRAP} of the Appendix; full line: set 2 given in Table \ref{table:S_STIRAPopt} of the Appendix. Panel (b) : Variation of the amplitudes by a common factor. $\mathcal{E}_{0,ref}$ designates the amplitudes of set 2 in Table \ref{table:S_STIRAPopt}.}
\label{fig_yield}
\end{figure}

We now explore another strategy inspired from the S-STIRAP scheme by involving a chain of intermediate states as in the $\pi $-pulse sequence. Only two supplementary transitions 1 – 2 and 2 – 3 are added to select 11 active states. The corresponding frequencies and dipoles are given in Table III of the Appendix. The pulses inducing the extreme transitions $56 - 48$ and 1 – 0 are driven in counterintuitive order and they are straddled by the nine pulses connecting the intermediate states. The first set of parameters is given in Table \ref{table:S-STIRAP} of the Appendix. We fix the maximum field amplitude as in the $\pi $ sequence and we increase the total duration of the simulation ${{t}_{\max }}$ by keeping the relative gap between the pulse positions $t_{c,ij}$ and the relative widths $\tau_{ij}$. As the transfer occurs in the ground electronic state, the aim is not to completely avoid population in the intermediate states since the radiative lifetimes are very long. We focus on the final yield in the ground $v$ = 0 state. The final population in the ground vibrational state as a function of ${{t}_{\max }}$ is drawn in dashed line in Figure \ref{fig_yield}(a). It increases in average but strongly fluctuates and exceeds 50$\%$ for a long ${{t}_{\max }}$ of the order of 1 $\mu$s. The yield could increase with a longer duration but it seems difficult to reach 100$\%$. 

In order to reduce the total pulse duration, we choose a shorter ${{t}_{\max }} = $700 ns corresponding to a local yield maximum of 40$\%$ and we slightly modify the pulse parameters, ${{t}_{c}}$, $\tau $ and ${{\mathcal{E}}_{0}}$. The parameters leading to 80$\%$ are given in Table \ref{table:S_STIRAPopt} of the Appendix. The evolution of the yield with ${{t}_{\max }}$ by using this parameter set is shown in full line in Figure \ref{fig_yield}(a). We also analyze the influence of the maximum amplitude on the yield. Figure \ref{fig_yield}(b) shows the effect of varying all the amplitude by a commen factor. The population in $v$ = 0 is drawn as a function of $\epsilon_0 / \epsilon_{0,ref}$ where $\epsilon_{0,ref}$ is given in Table \ref{table:S_STIRAPopt} of the Appendix.

\begin{figure}
\centering
\includegraphics[width=8cm]{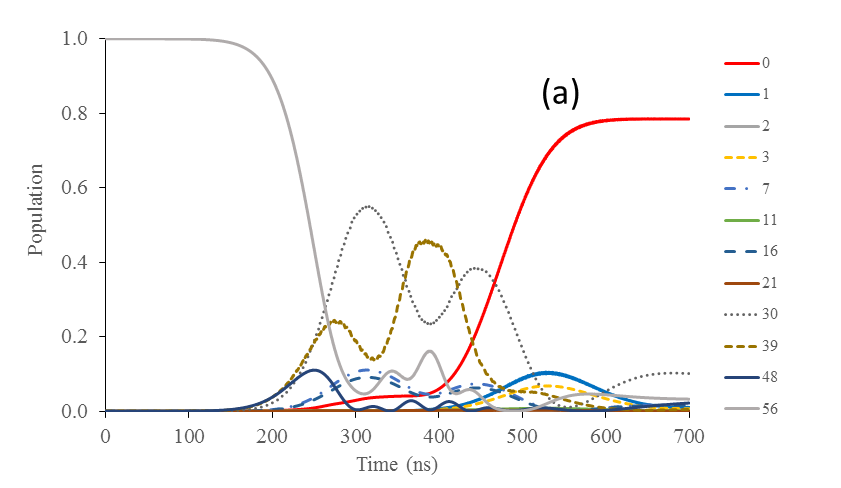}
\includegraphics[width=8cm]{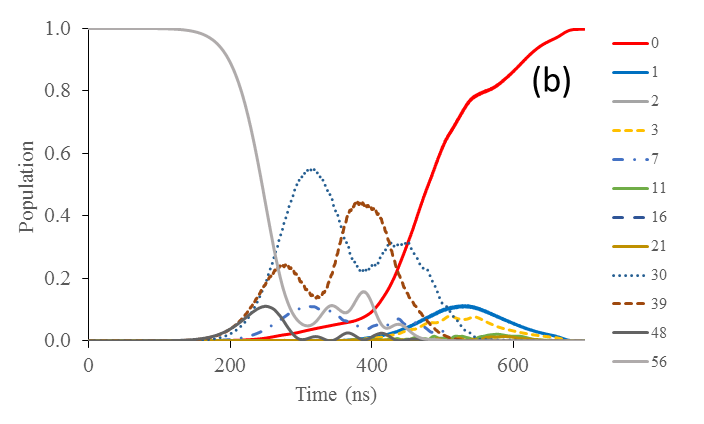}	
\caption{Panel (a): Population evolution driven by the S-STIRAP sequence with parameters given in Table \ref{table:S_STIRAPopt} of the Appendix with $t_{max}$ = 700 ns; panel (b): Evolution with the optimal field when the guess field is this S-STIRAP sequence. Only the states of the selected chain are drawn for clarity. }
\label{fig_pop_sti_oct}
\end{figure}

\begin{figure}
\centering
\includegraphics[width=8cm]{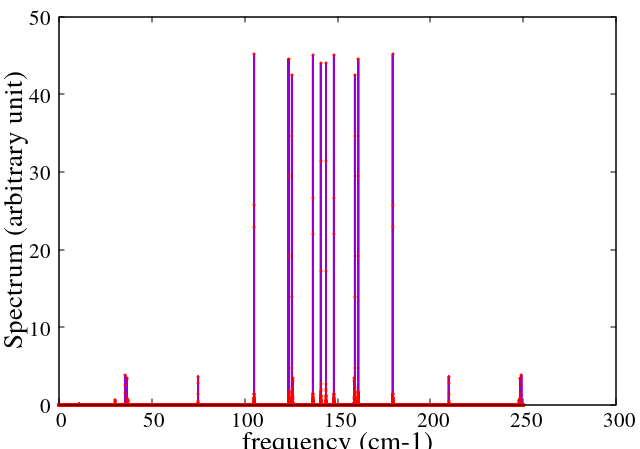}
\caption{Modulus of the Fourier transform of guess and OCT fields. Blue full lines: field of the S-STIRAP sequence with parameters given in Table \ref{table:S_STIRAPopt} of the Appendix with $t_{max}$ = 700 ns;, red dots: optimal field when the guess field is this S-STIRAP sequence.}
\label{fig_spectre_stoct}
\end{figure}

Finally, starting again from the new reference parameter set giving a yield 80$\%$ for ${{t}_{\max }}$ = 700 ns (Table \ref{table:S_STIRAPopt} in the Appendix), we optimize the field by OCT with this S-STIRAP field as guess. The algorithm converges in two iterations with a performance index of 100$\%$. This indicates the quality of the S-STIRAP field as guess field for the OCT. The populations driven by the S-STIRAP field and by the optimal field are compared in Figure \ref{fig_pop_sti_oct}. The profiles are very similar excepted at the end to ensure a yield of 100$\%$. Only the states of the selected chain are drawn for clarity but dynamics are performed in the complete basis set and the population in the other states remains negligible. The modulus of the Fourier transform of both fields are compared in Figure \ref{fig_spectre_stoct} revealing very few modification by the OCT. 

\subsubsection{Single optimal chirped pulse}
\begin{figure}
\centering
\includegraphics[width=8cm]{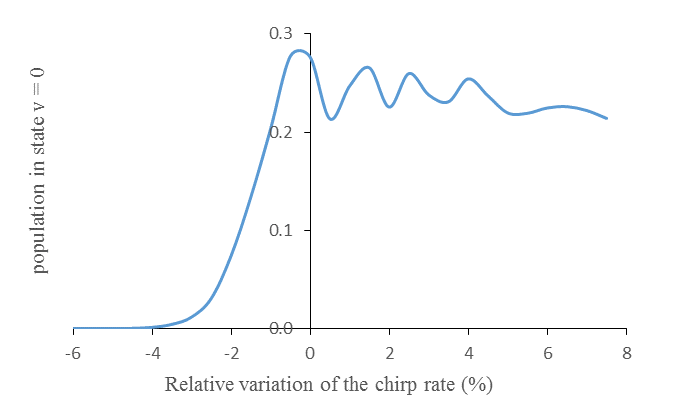}
\caption{Variation of the population in the vibrational ground state $v$ = 0 as a function of the relative variation of the chirp rate [Eq.(\ref{chirp})] with ${{\omega }_{0}}$= 2 cm$^{-1}$, ${{\omega }_{f}}$=20 cm$^{-1}$ and  ${{t}_{\max }}$ = 2.41 ns giving a rate  7.5 cm$^{-1}$/ns .}
\label{fig_chirprate}
\end{figure}

The last strategy exploits a chirped pulse \cite{Chelkowski1990} in order to induce the descent through the successive states that are coupled by couplings ${{\mu }_{v+1,v}}$ presenting a maximum around $v$ = 30 as illustrated in Figure\ref{fig_dip}(b). The energy gap between neighboring states is varying from 2.14 to 37.3 cm$^{-1}$. In order to find a guess field, we use a chirped pulse with constant amplitude $\mathcal{E} (t)={{\mathcal{E}}_{0}}\cos (\omega (t)t)$  where
\begin{equation}  
 \omega (t)={{\omega }_{0}}+\alpha t.
 \label{chirp}
\end{equation}
$\alpha$ being the chirp rate as defined in Eq.(\ref{chirp_rate}).
Finding a relevant parameter set is not easy and we give here an example providing a yield of 27$\%$ with ${{\omega }_{0}}$= 2 cm$^{-1}$, ${{\omega }_{f}}$=20 cm$^{-1}$ and  ${{t}_{\max }}$ = 2.41 ns.  The rate of frequency increase is 7.5 cm$^{-1}$/ns and this seems reasonable since it does not exceed 10$\%$ per ps. A field amplitude ${{\mathcal{E}}_{0}}$=$ 4 \times 10^{-5}$ a.u. slightly higher than in the previous strategy is necessary to get a non-negligible yield.  Figure \ref{fig_chirprate} shows the evolution of the yield with the rate of the chirp for the chosen amplitude and  ${{t}_{\max }}$.  One may see that the yield should strongly decreases if the rate decreases by 3$\%$. On the other hand, increasing the rate does not improve the yield.  

\begin{figure}
\centering
\includegraphics[width=8cm]{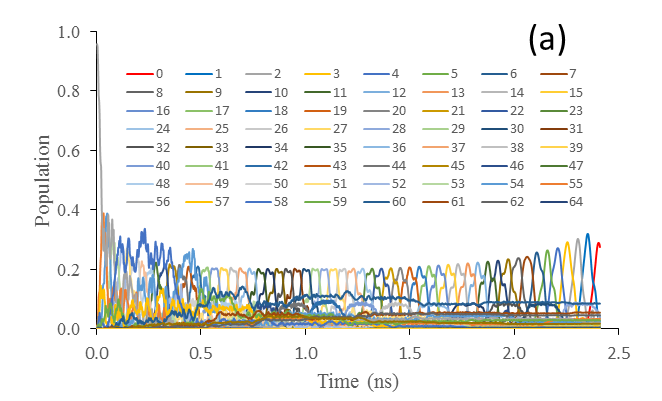}
\includegraphics[width=8cm]{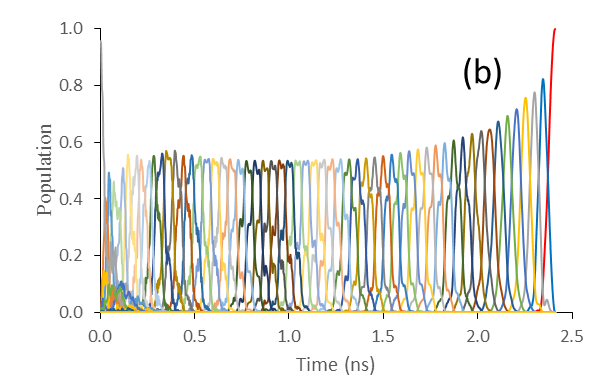}	
\caption{Panel (a): Population evolution driven by the chirp pulse [Eq.(\ref{chirp})] with parameters ${{\omega }_{0}}$= 2 cm$^{-1}$, ${{\omega }_{f}}$=20 cm$^{-1}$ and  ${{t}_{\max }}$ = 2.41 ns; panel (b): Evolution with the optimal field when the guess field is this chirped pulse.}
\label{fig_pop_chirp_oct}
\end{figure}

Figure \ref{fig_pop_chirp_oct}(a) shows the population evolution. Early dynamics are complicated and some states above $v$ = 56 are transitory populated. After 0.5 ns, one clearly observe the successive jumps towards the vibrational ground state. As in the previous strategy, we take this chirped pulse as a guess field for OCT. The optimal field again converges in very few iterations by confirming the efficiency of the guess field. Note that a sine square envelope is added in the OCT algorithm to ensure a smooth ramp up and dawn. The populations are given in Figure \ref{fig_pop_chirp_oct}(b). The optimal field simplifies the early dynamics. After 0.25 ns, there only remains  the regular successive jumps towards the vibrational ground state reached with 100$\%$. 

\begin{figure}
\centering
\includegraphics[width=8cm] {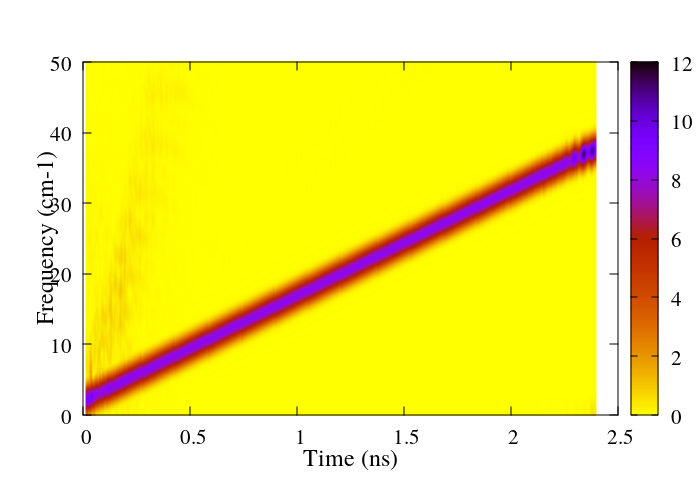}
\caption{Spectrogram in arbitrary units of the optimal field using the chirp as guess field. The chirp parameters are those of Figure \ref{fig_pop_chirp_oct}}
\label{fig_gabchirpoct}
\end{figure}

Figure \ref{fig_gabchirpoct} presents the time-dependent spectrogram of the optimal field obtained by Gabor transform using the Blackman window \cite{Fujimura94}. As expected from the population evolution shown in Figure \ref{fig_pop_chirp_oct}, OCT has added some frequencies during the early dynamics before 0.5 ns. The profile has some similarity with a second chirp. The other notable modification is the increase of amplitude at the end of the process. However the maximum amplitude does not exceed $5.8 \times 10^{-5}$ a.u. .

\section{Conclusion}
In this paper, several full optical strategies with their relative merits and limitations are studied for the formation of RbSr ultracold molecules in their absolute ground state.
Either two electronic states, or a single one are involved in the formation dynamics as controlled by linearly polarized laser pulses.
Loose of population in intermediate levels, efficiency and robustness for the chemical bond formation and the subsequent ro-vibrational transitions down to the rovibrational ground state are the main issues of the processes under consideration. 
The control schemes refer to two variants of STIRAP, to complete transfer optimized $\pi$-pulses, and to resonant excitation through chirped pulses.
The resulting control field characteristics basically involve peak intensities, pulse shape and duration, in given frequency domains, with the accompanying experimental feasibility criteria which are compared and discussed.

In summary, the methods, based on STIRAP and  involving two electronic states have the advantage of being already widely used in the ultracold literature.
In this context, we show how S-STIRAP is more appropriate than A-STIRAP for reducing the population in the intermediate loosely-bound level.
The caveat is that stronger fields  are required for the S version of STIRAP.

On the other side, methods involving  a single electronic state and referring to IR/THz pulses does not depend on the properties of excited states, but they must face the challenge of developing THz laser technology. The formation of chemical bond with THz pulses is demonstrated for chirped pulses and $\pi$-pulses. The adiabatic passage is more robust but the $\pi$-pulse require less intensity. This last point explain that in the near future, $\pi$-pulses would be probably more privileged. 

The exploration of the descent through the vibrational levels of the ground electronic state is a perspective since the pulse design in the micro-wave and far-IR range remains difficult nowadays. However, the simulations on this complex system confirm that such strategy will be conceivable with simple pulses of moderate intensity inspired from S-STIRAP and chirped pulses. Corrections found by OCT are very weak owing to very good guess fields. In the present investigation, we have used an optimization on a time grid but other methods based on the optimization of the pulse parameters, for instance by genetic algorithms should provide corrections more directly helpful for pulse shaping.  

	\section*{Acknowledgments}
This work has been supported in part by the project BLUESHIELD of \textit{Agence Nationale de la Recherche} (ANR-14-CE34-0006), and by the GDR 3575 THEMS of \textit{Centre National de la Recherche Scientifique} (CNRS). We acknowledge the use of the computing center MésoLUM of the LUMAT research federation (FR LUMAT 2764). 

\section{APPENDIX}
\label{appendix}

This Appendix displays specific molecule and laser parameters which are referred to when using $\pi$-pulses in the two S-STIRAP sequences illustrated in Figures \ref{fig_yield} and \ref{fig_pop_sti_oct} of the main text. Table \ref{table:pi_pulse} gives the frequency, the dipole transition, and the width of the Gaussian $\pi$-pulses. Note that the transitions $1 \rightarrow 2$ and $2 \rightarrow 3$ are used only in the STIRAP scheme. The first set of parameters used in the S-STIRAP sequence are gathered in Table \ref{table:S-STIRAP}. The parameters providing a yield of 80$\%$ are given in Table \ref{table:S_STIRAPopt}. They are obtained by systematically varying the parameters of the set given in Table \ref{table:S-STIRAP}.

\begin{table}[H]
\caption{Parameters of the transitions and widths of the $\pi$-pulses.} 
 			
	{\begin{tabular}{cccc} 
			 Transition    & $\omega_{ij}$ (cm$^{-1}$)   & Dipole $\mu_{ij}$ (a.u.) & $\tau_{ij}$ (ns) \\[0.5ex]
			\hline \noalign{\vskip 1.0ex}
			$56 - 48 $  & $30.5 $  & $1.446\times10^{-2} $  & $0.295$ \\[0.5ex]
			$48 - 39 $  & $75.1 $  & $2.267\times10^{-3} $  & $1.884$  \\[0.5ex]
			$39 - 30 $  & $125.9 $  & $1.563\times10^{-3} $  & $2.732$  \\[0.5ex]
			$30 - 21 $  & $178.7 $  & $7.305\times10^{-4} $  & $5.847$  \\[0.5ex]
			$21 - 16 $  & $123.7 $  & $1.807\times10^{-3} $  & $2.364$  \\[0.5ex]
			$16 - 11 $  & $141.0 $  & $1.196\times10^{-3} $  & $3.573$  \\[0.5ex]
			$11 - 7 $  & $125.5 $  & $1.096\times10^{-3} $  & $3.897$  \\[0.5ex]
			$7 - 3 $  & $136.9 $  & $4.485\times10^{-4} $  & $9.525$  \\[0.5ex]
			$3 - 1 $  & $72.6 $  & $1.749\times10^{-3} $  & $2.442$  \\[0.5ex]
			$1 - 0 $  & $37.3 $  & $2.767\times10^{-2} $  & $0.154$  \\[0.5ex]
			$2 - 1 $  & $36.6 $  & $3.925\times10^{-2} $  &  \\[0.5ex]
			$3 - 2 $  & $35.9 $  & $4.818\times10^{-2} $  &   \\[0.5ex]
		\hline
		\end{tabular}}
		\label{table:pi_pulse}
	\end{table}

\begin{table}[H]
\caption{First parameter set of the S-STIRAP sequence.} 
 			
	{\begin{tabular}{cccc} 
			 Transition    & ${{t}_{c}}/{{t}_{\max }}$    & $\tau /{{t}_{\max }}$ & ${{\mathcal{E}}_{0}}$(a.u.) \\[0.5ex]
			\hline \noalign{\vskip 1.0ex}
			$56 - 48 $  & $0.58 $  & $0.2 $  & $10^{-6}$ \\[0.5ex]
			$48 - 39 $  & $0.5 $  & $0.42 $  & $10^{-6}$  \\[0.5ex]
			$39 - 30 $  & $0.5 $  & $0.42 $  & $10^{-6}$  \\[0.5ex]
			$30 - 21 $  & $0.5 $  & $0.42 $  & $10^{-5}$  \\[0.5ex]
			$21 - 16 $  & $0.5 $  & $0.42 $  & $10^{-5}$  \\[0.5ex]
			$16 - 11 $  & $0.5 $  & $0.42 $  & $10^{-5}$  \\[0.5ex]
			$11 - 7 $  & $0.5 $  & $0.42 $  & $10^{-5}$  \\[0.5ex]
			$7 - 3 $  & $0.5 $  & $0.42 $  & $10^{-5}$  \\[0.5ex]
			$3 - 2 $  & $0.5 $  & $0.42 $  & $10^{-6}$  \\[0.5ex]
			$2 - 1 $  & $0.5 $  & $0.42 $  & $10^{-6}$  \\[0.5ex]
			$1 - 0 $  & $0.42 $  & $0.2 $  & $10^{-6}$  \\[0.5ex]
		\hline
		\end{tabular}}
		\label{table:S-STIRAP}
	\end{table}

\begin{table}[H]
\caption{Second parameter set of the S-STIRAP sequence.} 
 			
	{\begin{tabular}{cccc} 
			 Transition    & ${{t}_{c}}/{{t}_{\max }}$    & $\tau /{{t}_{\max }}$ & ${{\mathcal{E}}_{0}}$(a.u.) \\[0.5ex]
			\hline \noalign{\vskip 1.0ex}
			$56 - 48 $  & $0.58 $  & $0.2 $  & $ 3 \times10^{-7}$ \\[0.5ex]
			$48 - 39 $  & $0.38 $  & $0.42 $  & $10^{-6}$  \\[0.5ex]
			$39 - 30 $  & $0.38 $  & $0.42 $  & $10^{-6}$  \\[0.5ex]
			$30 - 21 $  & $0.5 $  & $0.42 $  & $1.16 \times10^{-5}$  \\[0.5ex]
			$21 - 16 $  & $0.5 $  & $0.42 $  & $1.16 \times10^{-5}$  \\[0.5ex]
			$16 - 11 $  & $0.5 $  & $0.42 $  & $1.16 \times10^{-5}$  \\[0.5ex]
			$11 - 7 $  & $0.5 $  & $0.42 $  & $1.16 \times10^{-5}$  \\[0.5ex]
			$7 - 3 $  & $0.5 $  & $0.42 $  & $1.16 \times10^{-5}$  \\[0.5ex]
			$3 - 2 $  & $0.5 $  & $0.42 $  & $10^{-6}$  \\[0.5ex]
			$2 - 1 $  & $0.5 $  & $0.42 $  & $10^{-6}$  \\[0.5ex]
			$1 - 0 $  & $0.4 $  & $0.2 $  & $ 3 \times10^{-6}$  \\[0.5ex]
		\hline
		\end{tabular}}
		\label{table:S_STIRAPopt}
	\end{table}

\bibliographystyle{unsrt}
\bibliography{Last_Paper}
\end{document}